\documentclass[12pt]{iopart}
\usepackage{graphicx}
\newcommand{\be}{\begin{eqnarray}}
\newcommand{\ee}{\end{eqnarray}}
\def\ben{\begin{equation}}
\def\een{\end{equation}}
\def\bena{\begin{eqnarray}}
\def\eena{\end{eqnarray}}
\def\non{\nonumber}

\renewcommand{\v}{\rm{v}}

\begin{document}

\title{Cosmic Acceleration Driven by Mirage Inhomogeneities}

\author{Christophe Galfard}

\address{D.A.M.T.P., Centre
for Mathematical Sciences, University of Cambridge, Wilberforce
road, Cambridge CB3 0WA, England} \ead{C.Galfard@damtp.cam.ac.uk}

\author{Cristiano Germani}

\address{D.A.M.T.P., Centre
for Mathematical Sciences, University of Cambridge, Wilberforce
road, Cambridge CB3 0WA, England} \ead{C.Germani@damtp.cam.ac.uk}

\author{Alex Kehagias}

\address{Physics Division, National Technical University of Athens, \\
15780 Zografou Campus,  Athens, Greece}
\ead{kehagias@central.ntua.gr}

\begin{abstract}
A cosmological model based on an inhomogeneous D3-brane moving in an
$AdS_5\times S_5$ bulk is introduced. Although there is no special
points in the bulk, the brane Universe has a center and is isotropic
around it.
The model has an accelerating
expansion and its effective cosmological constant is inversely
proportional to the distance from the center, giving a possible
geometrical origin for the smallness of a present-day cosmological
constant. Besides, if our model is considered as an alternative of early time acceleration, it is shown that the early stage
accelerating phase ends in a dust dominated FRW homogeneous
Universe. Mirage-driven acceleration thus provides a dark matter
component for the brane Universe final state. We finally show that
the model fulfills the current constraints on inhomogeneities.
\end{abstract}

\begin{flushright}
DAMTP-2005-87
\end{flushright}

\section{Introduction}

The motion of a D-brane in some gravitational  background follows  a
classical geodesic, so its induced metric is time dependent. The
motion of the brane leads to a time-dependent geometry, which is
interpreted as a cosmological evolution by a brane-observer. This
has been called mirage cosmology~\cite{KK} and  technically, there
are two steps in determining the resulting cosmological evolution .
To start with, the brane motion is determined by solving the
world-volume field equations, as follow from the Dirac-Born-Infield
(DBI) action, for the scalar fields which describe  the position of
the brane in the bulk. Then, the induced metric on the brane is
specified, now becoming an implicit function  of time, giving a
cosmological evolution for the brane. This cosmological evolution
can be reinterpreted in terms of cosmological ``mirage" energy
densities on the brane via a Friedman-like equation.

Various combinations of probe branes on simple backgrounds have been
analyzed~\cite{KK}, such as
stacks of Dp-branes at and out of extremality (black
Dp-branes), with and without additional constant antisymmetric
background tensors. They provide a cosmological evolution on the
probe brane which can be simulated by various types of mirage matter
on the brane. There is no a priori reason for the mirage
energy-momentum  tensor
to satisfy the energy conditions as it does not describe real matter
on the brane. What drives the
cosmological evolution is not  some form of matter or energy, but
the brane motion in the background geometry. Indeed,  in general, for a Dp-brane moving in the background of other Dp$'$-branes,
expansion on the world-volume of
the probe brane was found to correspond to an equation of state for the mirage
pressure and energy of the form $p=w \rho$ with $w>1$, violated by
all forms of matter.
It should be stressed however that  there are many exotic types of
mirage matter including $w$ values that are outside the range
$|w|\leq 1$ required by four-dimensional causality.

Another peculiarity of this framework is that individual densities
of dilute mirage matter can be negative (without spoiling the
overall positivity at late times). This is linked to the fact that
in this type of cosmology, the initial singularity is an artifact of
the low energy description.
This can be seen, for example, by studying brane motion in simple
spaces like $AdS_5\times S^5$ which are globally non-singular.
The induced cosmological evolution of a brane moving in such a space
has a typical expansion profile due to radiation and an initial
singularity (from the four-dimensional point of view). At  the
initial singularity the brane Universe joins a collection of
parallel similar branes and there is (non-abelian) symmetry
enhancement. The effective field theory breaks down and this gives
rise to the singularity.

The next obvious question is how ``real" matter/energy densities on
the brane affect its geodesic motion and the induced cosmological
evolution. This has been studied in the mirage-cosmology context  by
turning on electromagnetic energy on the brane. An important point
here is that the cosmological evolution is not driven by
four-dimensional gravity on the brane.

Motivated by the homogeneity of the D-brane backgrounds, (see
however \cite{horowitz}), a silent feature of the mirage-cosmology
scenario is that the resulting brane-cosmology is isotropic and
homogeneous. This is due to the fact that the scalars which
describe the position of the brane in the bulk geometry are taken to
only be time-dependent. This corresponds to a rigidly moving
three-brane in the bulk, and to our knowledge, on such backgrounds,
the mirage matter alone cannot account for the acceleration of the
Universe's expansion \cite{tetradis}\footnote{In order to obtain an accelerating Universe in this scenario,
it is indeed necessary to turn on a tachyonic bulk field. In this case it is also possible to have a mechanism for
the reheating of the Universe \cite{papa}.}. We here consider a  brane moving in such a way
that its motion depends not only on time but also on some other
space directions. In this case, the induced metric on the brane is
neither isotropic nor homogeneous. %cr
Inhomogeneous cosmologies are well studied in the literature, and it
has been shown that the current data cannot rule them out (see
\cite{kra} for a review). Moreover, it has been suggested in
\cite{notari} that late time cosmology cannot be approximately
homogeneous. However, the recent proposals that inhomogeneities in a
matter dominated Universe, on either
super-horizon~\cite{Kolb:2005da}  or sub-horizon
scales~\cite{rasanen} may influence the expansion rate at late times
have been
criticized~\cite{rasanen},\cite{Geshnizjani:2005ce},\cite{siegel},\cite{giovani}.
In fact, it has been shown that sub-horizon~\cite{siegel} and
super-horizon~\cite{rasanen}
perturbations,\cite{Geshnizjani:2005ce},\cite{giovani} are not a
viable candidate for explaining the accelerated expansion of the
Universe. This is simply because as long as energy conditions are
valid, inhomogeneous models can never lead to accelerated
expansion~\cite{giovani}. It should be noted that inhomogeneous
cosmological models in the context of string theory have previously
been discussed~\cite{veneziano}.

Here, however, we will show that in mirage cosmology,
inhomogeneities may indeed lead to accelerated cosmic expansion
either at early or late times. In particular, inhomogeneities
compatible with $SO(3)$ symmetry may lead to an accelerating
expansion phase for a probe D3-brane moving in the near horizon
limit of a background of a stack of D3-branes. When this phase
happens at early times, it may  provide a mechanism for inflation.
In that case, exit of the inflation is achieved when inhomogeneities
are diluted enough due to inflation. We will see that what remains
of these inhomogeneities is similar to pressureless matter. Thus, in
this scenario, mirage matter left over at the end of the
inflationary phase, may be a dark matter component. On the other
hand, when the accelerating phase is at late times, mirage cosmology
can provide an alternative to $\Lambda$CDM.

It should also be noted that our model does not have any initial
singularity so that it naturally fits the homogeneity of the CMB
spectra. In fact, in the absence of big bang-like singularities, all
parts of the universe may have been in causal contact in the far
past. Moreover we also show that the observed small scale
homogeneities (in the range of $100\ \div 140\ \mbox{Mpc}$
\cite{Hogg}) can be implemented in our model and related to the best
fit cosmological constant of \cite{team}.

\section{Set-up}

In general, a probe D-brane light enough for its back-reaction to be
neglected  will follow  a geodesic as it moves in some string-theory
background. Due to its motion, the induced world-volume metric on
the probe brane becomes a function of time so that, from the brane
``residents'' point of view, a changing (expanding or contracting)
Universe is experienced. This is the basic idea behind mirage
cosmology. The simplest case corresponds to a D3-brane moving in a
static spherically symmetric gravitational background with a dilaton
as well as a RR background $C(r)=C_{0...3}(r)$ with a self-dual
field strength.  The probe brane will in general start moving in
this background along a geodesic and its dynamics is governed by the
DBI action. In the case of maximal supersymmetry the DBI action is
\begin{eqnarray} S=T_3\!\int\! d^{4}\xi
e^{-\phi}\sqrt{-det(\hat G_{\alpha\beta} +(2\pi
\alpha')F_{\alpha\beta}-B_{\alpha\beta})}+ T_3\!\int\! d^{4}\xi
~\hat C_4 +{\rm anomaly~ terms}\, , \label{action}
\end{eqnarray}
where world-volume fermions have been ignored and $\xi^\alpha
~(\alpha=0,1,2,3)$ parameterize the D3-brane world-volume. The
embedded data are given by \be
 \hat G_{\alpha\beta}=G_{\mu\nu}{\partial x^{\mu}\over
\partial\xi^{\alpha}}{\partial x^{\nu}\over
\partial\xi^{\beta}}\, ,
\ee etc. For simplicity we will assume that the  $B$-field is zero
and there is no electromagnetic field on the brane.

In particular, we will consider here a probe D3-brane moving in the
near-horizon limit of a stack of D3-branes with metric
\begin{eqnarray}
ds^2=\frac{r^2}{L^2}\left(-dt^2+d\rho^2+\rho^2d\Omega_2^2\right)+\frac{L^2}{r^2}dr^2+L^2d\Omega^2_5
\ ,
\end{eqnarray}
where we have chosen spherical coordinates for the brane
directions. This is just an $AdS_5\times S^5$ space-time and $L$
is the $AdS_5$ length defined by $L^4=4\pi g_s N$, where $g_s$ is the string coupling and $N$ the D3-brane charge. The
 RR field is
 \be
 C_{0..3}=\frac{r^4}{L^4}\ee
 and the dilaton $\phi$ is constant. We will regard
our Universe as a D3-brane moving in this background.

We now consider an inhomogeneous trajectory for the brane motion
$r=r(t,\rho)$. The case of homogeneous trajectory has been studied
before \cite{KK}. No matter is included on the brane and only
geometrical quantities related to the chosen embedding will play a
role. The induced metric is then
\be
ds^2_{ind}=-\left(\frac{r^2}{L^2}-\frac{\dot r^2
L^2}{r^2}\right)dt^2+\left(\frac{r^2}{L^2}+ \frac{r'^2
L^2}{r^2}\right)d\rho^2+2\frac{\dot r r'}{r^2}L^2d\rho\
dt+\frac{r^2}{L^2}\rho^2d\Omega^2_2\ , \label{induced}
\ee
where $r=r(t,\rho)$ and a dot denotes derivative with respect to
time $(~\dot{} =\partial_t)$, whereas a prime denotes derivative
with respect to brane-radial distance $(~'=\partial_\rho)$.

Due to re-parametrization invariance, there is a gauge
freedom, which may be fixed by choosing the static gauge,
$x^{\alpha}=\xi^{\alpha}$, so that in this gauge, the total
Lagrangian is
 \be \frac{{\cal L}}{4\pi
T_3}=\frac{r^4}{L^4}\rho^2\sqrt{1+\frac{r'^2 L^4}{r^4}-\frac{\dot
r^2 L^4}{r^4}}-\frac{r^4}{L^4}\ . \ee In the following we will
consider the ``non-relativistic'' approximations \be \label{approx1}
\frac{r'^2 L^4}{r^4}\ll 1\ ,\ \frac{\dot r^2 L^4}{r^4}\ll 1\,  .
\label{non-rel} \ee In this case the field equations $\delta {\cal
L}=0$ reduce to \be \ddot r-\frac{1}{\rho^2}\partial_\rho
\left(\rho^2 r'\right)=0\ . \ee As the bulk is regular, the only
possible brane singularity happens when the brane shrinks to a
point, i.e. when $r=0$. The most general solution without
singularity is \be r=\frac{\mu}{2}(t-t_0)^2+\frac{\mu}{6}\rho^2+r_0\
, \label{soll} \ee as one can check by calculating all the curvature
invariants. Here $r_0 (> 0)$ corresponds to the position of the
center of the brane at some fiducial time $t_0$ and $\mu$ is a
mass-parameter. We also would like to stress that this solution does
not have a homogeneous limit, and thus belongs to a new branch of
solutions.

Note that the above discussion is valid in the non-relativistic
approximation (\ref{non-rel}). Using (\ref{soll}), it is easy to
verify that a sufficient condition for the validity of this
approximation is \be \label{non relativistic approximation} \mu\ll
\frac{r_0^3}{L^4}\, . \ee

\section{Accelerating Universe}

In this section, we will show that the expansion rate of geodesic
congruences is accelerated with a natural ending, the end state
corresponding to pressureless mirage matter. Depending on when
mirage effects dominate conventional Einstein gravity, inhomogeneous
mirage matter may lead to  either early or late time cosmological
acceleration. In particular, it may provide an alternative to
inflation, or to the late-time cosmic expansion we observe today. We
may employ the present model for inflationary cosmology if mirage
matter is dominating at early times. Then, not only our Universe is
accelerated at early time, but it does not have a cosmological
singularity, avoiding the existence of a particle horizon. In our
model, in fact, all parts of the Universe are connected to each
other by a time-like curve and are separated by a finite amount of
proper time. This does not happen for non-inflating singular
cosmologies like the FRW model. In our case, therefore, the
homogeneity of the CMB spectra is natural.

\subsection{Light-cone acceleration}

To simplify the problem, we shall approximate the induced metric
(\ref{induced}) by \be \label{metric} ds^2_{i}\simeq
\frac{r^2}{L^2}(d\eta^2)\ , \ee where $d\eta^2$ is the four
dimensional Minkowski metric. For our purpose, (\ref{metric}) will
be a good approximation to (\ref{induced}) if the relative
discrepancy of the Christoffel and the second variations of the
determinant of the metrics are very small compared to 1 \footnote{
  i.e. if 'i' corresponds to the induced metric (\ref{induced}) and
  'c' to the conformal one (\ref{metric}), we want
$\frac{ {}^{i} \Gamma^\alpha_{\beta
    \gamma}-{}^{c}\Gamma^\alpha_{\beta
  \gamma}}{{}^{i}\Gamma^\alpha_{\beta \gamma}} <<1$ and
    similarly, we require that $\frac{\partial^2
      \sqrt{-{}^{i}g}-\partial^2\sqrt{-{}^{c}g}}{\partial^2
      \sqrt{-{}^{i}g}}<<1$, where ${}^{i/c}g$
    is the determinant of the corresponding metric}.
One can check that these conditions are met assuming the
non-relativistic approximation (\ref{non relativistic
approximation}). The four-velocity of a null radial geodesic in the
metric (\ref{metric}) is then easily found to be
\begin{eqnarray}\label{four}
k^\alpha=\left[-\frac{F(\rho+t)}{(\mu\rho^2+3\mu(t-t_0)^2+6
r_0)^2},\frac{F(\rho+t)}{(\mu\rho^2+3\mu(t-t_0)^2+6 r_0)^2},0,0\right]\ ,
\end{eqnarray} where $F(\rho+t)$ is a non-determined function that
depends on the parametrization chosen for the null geodesic.

Defining the expansion scalar as $\theta=\nabla_\alpha k^\alpha$,
the acceleration parameter $Q_n$ of a null-congruence is \cite{HE}
\begin{eqnarray}
Q_n=k^\alpha\nabla_\alpha \theta+\frac{1}{2}\theta^2=96\mu^2
F(\rho+t)^2\frac{(t-t_0)\left[\rho-(t-t_0)+\frac{r_0}{\mu(t-t_0)}\right]}{\left[\rho^2+3(t-t_0)^2+6r_0
\right]^6} \label{ra} .
\end{eqnarray}
We recall that for a null geodesic vector $k^\alpha$,  the following
geometrical condition holds
 \be
 k^\alpha\nabla_\alpha \theta+\frac{1}{2}\theta^2=-R_{\alpha\beta}k^\alpha
k^\beta+\omega^2-\sigma^2\ , \ee where the vorticity
$\omega_{\alpha\beta}$ and shear $\sigma_{\alpha_\beta}$ are defined
as \be \omega_{\alpha\beta}=\nabla_{[\alpha}k_{\beta]}\, , ~~~~
\sigma_{\alpha\beta}=\tilde h^\mu_{(\alpha}\tilde
h^\nu_{\beta)}\nabla_\mu k_\nu-\frac{1}{2}\theta^2\, , \ee
 in terms of the two-metric $\tilde h_{\alpha\beta}$,
which satisfies  $\tilde h_{\alpha\beta}k^\alpha=0$. One can show
that the four velocity (\ref{four}) has zero vorticity and shear for
the lowest order induced metric (\ref{metric}). So in order to have
acceleration ($Q_n>0$), the null energy conditions $
R_{\alpha\beta}k^\alpha k^\beta\geq 0$ must be violated. Indeed, for
the metric (\ref{metric}), $R_{\alpha\beta}k^\alpha k^\beta=-Q_n$
so that when the Universe's light-cone expansion accelerates, the
null energy conditions are violated. However, since no physical
matter has really been used, there is no violation of any
fundamental physical law. There is no reason to expect mirage matter
to satisfy any energy condition, since it does not corresponds to
any form of real matter. This has also been stressed
before~\cite{KK}, where mirage matter violating causality bounds was
found to drive cosmological expansion.

It is now easy to see that the light cone is accelerating ($Q_n>0$)
for $\rho+\frac{r_0}{\mu(t-t_0)}>(t-t_0)$ and that the acceleration
naturally ends when $\rho+\frac{r_0}{\mu(t-t_0)}=t-t_0$. Thus, the
larger the radial distance, the longer the
 accelerating era. %cr

We would like to stress that the acceleration is only indirectly due
to inhomogeneities. In fact, as we showed, the only contribution to
the Raychaudhuri equation (\ref{ra}) is coming from the negative
mirage energy density of the brane inhomogeneities.

\subsection{Time-like congruences}

For a time-like congruence, the geodesic equation is
$u^\alpha\nabla_\alpha u^\beta=0$, where the vector $u^\alpha$ is
normalized \be \label{norm} u^\alpha u_\alpha=-1\ . \ee For a radial
motion, we have that
$u^\alpha=\Big{(}u^t(t,\rho),u^\rho(t,\rho),0,0\Big{)}$ and Eq.
(\ref{norm}) is satisfied for \be
u^t=-\sqrt{\frac{L^2}{r^2}+(u^\rho)^2}\ . \ee As it is difficult to
solve the above geodesic equation, some approximations must be used
to get analytical results. Let us consider the following
quasi-homogeneous congruence \be\label{app} u^\rho=\v \ll L/r\ ,\ee
so that the time component of $u^\alpha$ can be approximated by \be
u^t=-\frac{L}{r}\left[1+\frac{r^2}{2 L^2} \v^2+{\cal
O}\left(\frac{r^4}{L^4}{\v^4}\right)\right]\, . \ee The geodesic
motion turns out to be given, to leading order, by a single equation
for $u^\rho$ \be \dot{\v}=\frac{r'L}{r^2}\ , \ee and its solution
with initial condition $\v(t=t_0,\rho)=0$ is \be \v=\frac{\alpha
(t-t_0)}{2\beta\left[(t-t_0)^2+\beta\right]}+\frac{\alpha}{2\beta^{3/2}}
\arctan\left(\frac{t-t_0}{\beta^{1/2}}\right)\ . \ee The parameters
$\alpha,\beta$ are given by \be\label{alpha and beta}
\alpha=\frac{4\rho L}{3\mu}\ , \ \beta=\frac{\rho^2}{3}+\frac{2
r_0}{\mu}\ , \ee in terms of which we may express the solution in
Eq.(\ref{soll}) as \be r=\frac{\mu}{2}\left((t-t_0)^2+\beta\right)\
. \ee

Since, to us, the Universe looks homogeneous, isotropic and
accelerated,  we may also consider our position as being very far
from its center. According to the condition (\ref{non relativistic
approximation}), we can consider $\mu$ small and/or $r_0$ large so
that $\beta$ is large. In the $\beta\gg (t-t_0)^2$ limit we have \be
\v=\frac{\alpha(t-t_0)}{\beta^2}+{\cal
O}\left(\frac{1}{\beta^3}\right)\ , \ee and the condition
(\ref{app}) is fulfilled for \be (t-t_0)\ll \frac{\rho}{2}+\frac{3
r_0}{\mu\rho}\ . \ee Considering units such that our position is at
$\rho>1$ and assuming $t/t_0>1$, we find that a sufficient condition
for the validity of our approximation actually is \be \beta\gg
(t-t_0)^2\ . \ee We may now calculate the acceleration parameter
$Q_t$. For time-like geodesics, it is
 $$Q_t=\dot \theta+\frac{1}{3}\theta^2\, . $$
In the large $\beta$ limit, we find that $Q_t$ is given by \be
Q_t= \frac{32L^2}{3\mu^2\beta^3}+{\cal{O}}(\frac{1}{\beta^4})\, .
\ee As for null geodesics, we find that these time-like geodesic
congruences are accelerating ($Q_t>0$). Considering our position
in the Universe to be $\rho=\rho_0\gg \sqrt{6 r_0/\mu}$, we have
\be Q_t\simeq \frac{288 L^2}{\mu^2\rho_0^6}=\Lambda\ , \ee where
$\Lambda$ is the cosmological constant as measured today. Thus,
our relative position with respect to the center is
\be\label{Lambda} \rho_0\simeq \left(\frac{288
L^2}{\mu^2\Lambda}\right)^{1/6}\ , \ee and therefore, the measured
small cosmological constant is in principle compatible with a
large distance of our position with respect to the center of the
Universe. It should be stress that a fine-tuning of $Q_t$ is needed in order to match the observed acceleration.
However, in this case the smallness of the latter is geometrical
and due to the large distance of our position from the center. Still there is a fine-tuning as
we have to choose the parameters $\mu_0,L$, which nevertheless
is not better than the usual fine-tuning of the cosmological constant. Clearly, the
present  model is not relevant for
 the cosmological constant problem arising from the contribution of
supersymmetry breaking, phase transitions etc. in the vacuum energy. As usual, we assume that a mechanism
exists which neutralize all these vacuum-energy sources leaving only the mirage contribution to account for the observed acceleration.

\begin{figure}[t]
\centering
\includegraphics[angle=0,width=4in]{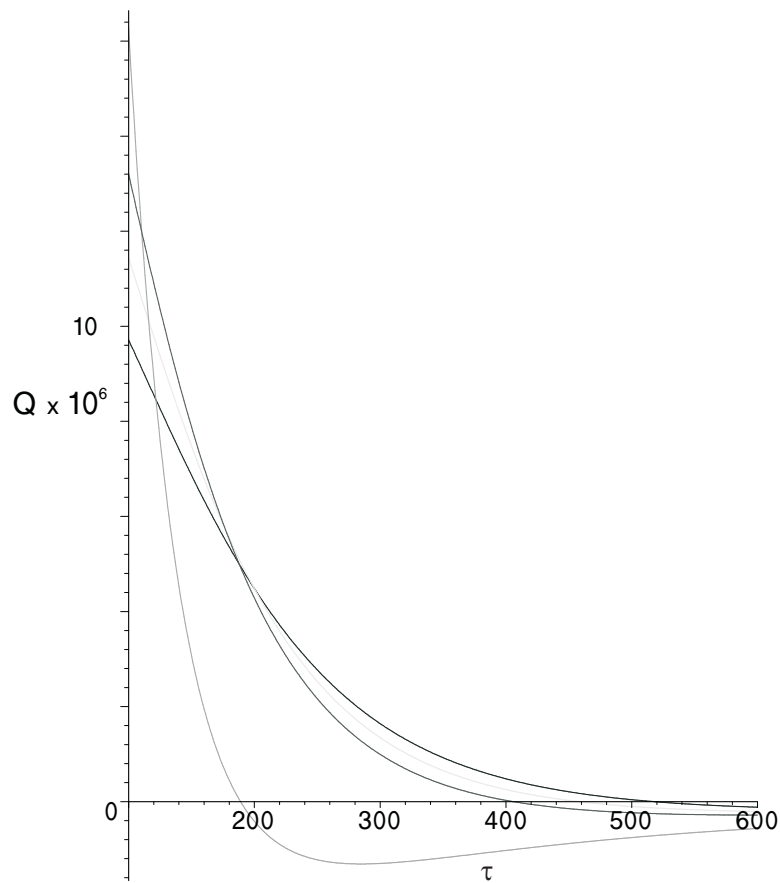}
\caption{``FRW-like'' expansion $Q=\ddot a/a$. Units of $AdS$
length, i.e. $L=1$. Decreasing curves close to $\tau=0$ correspond
to $\rho=10,200,220,240$. Values used, $t_0=0$, $r_0=1$ and
$\mu=10^{-4}$.\label{fig1}}
\end{figure}

One may also verify that the vorticity $\omega_{\alpha\beta}$ and
the shear  $\sigma_{\alpha \beta}$ given in terms of   the three
metric $h_{\alpha\beta}=g_{\alpha\beta}+u_{\alpha}u_\beta$ by \be
\omega_{\alpha\beta}=\nabla_{[\alpha}u_{\beta]}\, ,
~~~~\sigma_{\alpha\beta}=h^\mu_{(\alpha}h^\nu_{\beta)}\nabla_\mu
u_\nu-\frac{1}{3}\theta^2\,, \ee vanish to lower order and \be
\omega^2=\omega_{\alpha\beta}\omega^{\alpha\beta}={\cal{O}}\big{(}\frac{1}{\beta^4}\big{)},
~~~~\sigma^2=
\sigma_{\alpha\beta}\sigma^{\alpha\beta}={\cal{O}}\big{(}\frac{1}{\beta^4}\big{)}\,
. \ee In addition, a short calculation of the energy conditions
reveals that \be R_{\alpha\beta}u^\alpha
u^\beta=-\frac{32L^2}{3\mu^2\beta^3}+{\cal{O}}(\frac{1}{\beta^4})\,
, \ee and clearly, the strong energy conditions
$R_{\alpha\beta}u^\alpha u^\beta\geq 0$ are violated, while the
expansion accelerates. For the  time-like geodesic congruence we are
discussing,   the Raychaudhuri equation is satisfied, \be \dot
\theta+\frac{1}{3}\theta^2=-R_{\alpha\beta}u^\alpha
u^\beta+\omega^2-\sigma^2\ . \ee A comparison to the FRW homogeneous
model can easily be made by approximating the induced metric
(\ref{metric}) by \be ds^2\simeq
((t-t_0)^2/2+\rho_0^2/6+r_0)^2\left(-dt^2+d\varrho^2+\varrho^2
d\Omega_2^2\right)\ , \ee around some fixed value of the radial
distance $\rho=\rho_0$ with $\rho=\rho_0+\varrho$ and
$(t-t_0)\gg\varrho$ Then in ``synchronous time''
$d\tau=((t-t_0)^2/2+\rho_0^2/6+r_0) dt$ one can rewrite the metric
as a FRW-like one \be ds^2\simeq -
d\tau^2+a(\tau)^2\left(d\varrho^2+\varrho^2 d\Omega_2^2\right)\ .
\ee The expression for $a(\tau)$ is rather involved so instead of
giving
 its explicit form,  Fig.(\ref{fig1}) plots the
acceleration parameter $\frac{\partial_\tau ^2 a}{a}=\partial_\tau
\theta+1/3 \theta^2$, where $\theta$ is calculated for time-like
geodesic observers. At large proper time, the asymptotic expansion
of the acceleration is \be \frac{\ddot a}{a}\sim
-\frac{2}{9}\tau^{-2}\ , \ee which is the acceleration of a
dust-dominated Universe in a FRW background, as we shall now show.

Considered as a model for inflation, our model has certain
advantages. In particular, a
 mirage-driven inflationary phase
 naturally ends when inhomogeneities are diluted enough by the inflation. Indeed,
when $\mu t^2$ dominates $\mu \rho^2/6+r_0$ in $r$, the metric
becomes approximately homogeneous \be ds^2\simeq
\frac{\mu^2}{4L^2}t^4\left(-dt^2+dx_3^2\right)\ . \ee We can
consider $t$ as a conformal time and  define the synchronous time
$\tau=\mu t^3/6 L$ so that \be ds^2\simeq
-d\tau^2+\left(\frac{\tau}{\tau_0}\right)^{4/3}dx_3^2\ , \ee where
$\tau_0=3\sqrt{2L/\mu}$. At late time the spacetime looks like FRW
with a scale factor $a(\tau)\sim \tau^{2/3}$, as expected in a
dust-dominated Universe. Moreover, the inhomogeneous mirage
acceleration provides a dark matter component left over from an
early-time inflationary phase .

\section{Homogeneity bounds}

Mirage inhomogeneities  may also account for the observed
accelerating expansion of the Universe deduced form Type SNIa
supernova observations and CMB WMAP data \cite{team}. In this case,
it would be enough for us to be at a distance from the center such
that $\rho>t-t_0+\frac{r_0}{\mu (t-t_0)}$ to have acceleration of
the light cone, thus explaining the  observed cosmic acceleration.
However there are constraints coming from the homogeneity  bounds of
the Universe.

The first one is on the homogeneity of the CMB spectra. In standard
cosmology, an initial big bang-like singularity prevents very
distant objects from having been in causal contact in the far past.
In the present model, however, there is no initial singularity. It
is thus allowed to consider a homogeneous radiation put on top of an
inhomogeneous spacetime. Our model should be compared to tests on
homogeneity at small scales. In particular, the Luminous Red
Galaxies (LRG) data~\cite{Eisenstein:2001cq} of the Sloan Digital
Sky Survey~\cite{sdss} can be used as a measure of the homogeneity
of the Universe. The analysis
 suggests that the Universe is homogeneous for comoving distances
 of the order of $R\sim 70 h^{-1} \mbox{Mpc}$. It actually reveals
 that our universe is homogeneous from around
$100\ \mbox{Mpc}$ up to distances of $140\
\mbox{Mpc}$~\cite{Hogg},\cite{Yadav}. This means that within this
 shell, the number of
luminous objects in a given volume $V$ of a box of size $\ell$
scales like $\ell^3$. Possible inhomogeneities are thus restricted
and a comparison of the present model with observational data is
welcome. So far, the model considered was empty of real matter. We
consider the galaxies to be dust and would like to see that the
mirage inhomogeneities are not inconsistent with the observed
homogeneity of the Universe. In effect, since for dust the energy
contained in a given volume will grow with the volume, we can
roughly say that the number of object in the volume is proportional
to the volume itself. Therefore, our Universe will look like
homogeneous if the three-volume scales like the distance to the cube
up to $140\ \mbox{Mpc}$. For this, we need to write down the induced
metric (\ref{metric}) in comoving coordinates. This is not an easy
task but we may find an approximate answer. First we need to find
the coordinate transformation from $(t,\rho,\theta,\phi)$ to
$(\tau,R,\theta,\phi)$ such that the 4-velocity
$u^\alpha=-\delta^\alpha_\tau$ where $\tau$ is the proper time. So
we impose:
 \be\label{first}
-1=u^\tau=\dot\tau u^t+\tau' u^\rho\ , \ee and \be\label{second}
0=u^R=\dot R u^t+R' u^\rho \ . \ee As before we will use the
large-$\beta$ approximation. In particular, since  we are interested
in our position in the Universe, we will consider $\beta \sim
\rho^2/3$ in (\ref{alpha and beta}).

Equations (\ref{first},\ref{second}) are solved by
\begin{eqnarray}
R=\frac{\mu}{3}(t-t_0)^2+\frac{\mu}{6}\rho^2 \simeq
\frac{\mu}{6}\rho^2  \, ,~~~~
\tau=\frac{2}{3L}(t-t_0)[R+\frac{\rho^2 \mu}{12}]\, ,
\end{eqnarray}
which can be inverted, giving
\begin{eqnarray}
\rho=\sqrt{\frac{6R}{\mu}}\, , ~~~~ t=L\frac{\tau}{R}+t_0 \ .
\end{eqnarray}
In $(\tau,R,\theta,\phi)$ coordinates, the metric is
\begin{eqnarray}
g_{TT}&=&-1\ ,\cr
g_{RR}&=&g_{\rho\rho}\left[\left(\partial_R\rho\right)^2-\left(\partial_R
t\right)^2\right] \simeq \frac{3}{2}\frac{R}{\mu L^2}\ ,\cr
g_{RT}&=&g_{tt}\left[\partial_R t\partial_\tau t-\partial_R
\rho\partial_\tau \rho\right] \simeq \frac{\tau}{R} \, ,
\end{eqnarray}
so that the line element reads \be \label{metric1}
ds^2=-d\tau^2+\frac{3}{2}\frac{R}{\mu
  L^2}dR^2+2\frac{\tau}{R}dRd\tau+\frac{6R^3}{\mu L^2}d\Omega^2_2\ .
\ee In order to calculate the three volume in a box around a given
point, we introduce a new radial variable $\xi$ so that the
circumference of the circles in $d{\Omega_2}^2$ is $2\pi\xi$ \be
\xi=\sqrt{\frac{6}{\mu L^2}}R^{3/2}\ , \ee in terms of which
(\ref{metric1}) becomes \be \label{metric2}
ds^2=-d\tau^2+\frac{1}{9}d\xi^2+\frac{4\tau}{3\xi}d\xi
d\tau+\xi^2d\Omega^2_2\ . \ee To calculate the volume of a box
centered at our position it is useful to use 'cartesian' coordinates
so that $d\xi^2+\xi^2 d\Omega^2_2=dx^2+dy^2+dz^2$, and
$\xi^2=x^2+y^2+z^2$. The line element becomes:
\begin{eqnarray}
ds^2&=&-d\tau^2+(1-\frac{8}{9}\frac{x^2}{\xi^2})dx^2+(1-\frac{8}{9}\frac{y^2}{\xi^2})dy^2+(1-\frac{8}{9}\frac{z^2}{\xi^2})dz^2+\frac{4}{3}\frac{x\tau}{\xi^2}dx
d\tau \non \\&&+\frac{4}{3}\frac{y\tau}{\xi^2}dx d\tau
+\frac{4}{3}\frac{z\tau}{\xi^2}dx d\tau
-\frac{16}{9}\frac{dxdy}{\xi^2}-\frac{16}{9}\frac{dxdz}{\xi^2}-\frac{16}{9}\frac{dydz}{\xi^2}\
.
\end{eqnarray}

For simplicity, we consider our spatial position in the Universe to
be at the point $(x_0,x_0,x_0)$, where, as discussed before, $x_0$
is large with respect to any other length considered. There is
nothing particular about this point and it has been chosen for
computational simplicity only. The volume of a box of coordinate
side size $2 x_1$, centered at $(x_0,x_0,x_0)$ can now be estimated.
We first need to find the proper length of the box. The proper
length of side of the box is given by \be \ell=\int_{-x_1}^{x_1}
v_\alpha dx^\alpha\ , \ee where the 4-vector $v_\alpha$ is tangent
to one of the box's edge and satisfies $v_\alpha u^\alpha=0$ and
$v_\alpha v^\alpha=1$.

The volume $V$ of the box for large $x_0$ is then: \be
\label{volume} V \propto
\ell^3\left(1-\frac{8}{\sqrt{33}}\frac{\ell}{x_0}\right)=\ell^3\left(1-\frac{4
\ell}{\sqrt{11}\sqrt{2}}\sqrt{\Lambda}\right)\ , \ee where, tracing
back the coordinate transformations, we used \be x_0=\frac{\mu
{\rho_0}^3}{6\sqrt{3}L}\simeq 2\sqrt{\frac{2}{3}}\Lambda^{-1/2}\ ,
\ee where we used (\ref{Lambda}) for the last step.

To compare with observations we take
$\Lambda^{-1/2}=2.6\times10^3~{\rm Mpc}$ \cite{team}. We check that
the fractal dimension of our Universe up to $140\ \mbox{Mpc}$ away
from the observer, and with a cut off at the scale of
inhomogeneities taken to be at $100\ \mbox{Mpc}$ \cite{Hogg}, is
within the observed homogeneity bound. Denoting the $100\
\mbox{Mpc}$ cut-off values for the volume and length by $V_c$ and
$l_c$ respectively, the fractal dimension $D$ we are looking for is
defined by: \be D=\frac{\log{\frac{V}{V_c}}}{\log{\frac{l}{l_c}}}
\ee which we find to be $D=3 \pm 0.03$, i.e. within the observed
bound \cite{Hogg}.

\section{Conclusions}

In this paper, we introduced an inhomogeneous type of solution of a
model of a D3-brane moving in an $AdS_5\times S_5$ bulk. We assumed
that all matter contributions are sub-dominant to the geometrical
mirage energy contribution. Our brane is inhomogeneous and although
from the bulk viewpoint there is no special point, a center of the
Universe must be introduced. This center appears due to the
inhomogeneity, it is not singular and the Universe is isotropic
around this point. The cosmological evolution does not have an
initial singularity and does not have a cosmological singularity.
This prevents us from having a horizon problem. We found that the
brane is initially accelerated with increasing magnitude as the
distance from the center decreases. The acceleration naturally ends
in a FRW-like dust dominated Universe at late time as the distance
from the center becomes large. The mirage matter can then be thought
of as a component of the observed dark matter.

We may also consider this model for explaining the present-day
observed cosmological acceleration. In this case, the cosmological
constant can be understood as being geometrical, since it is
inversely proportional to the distance to the center of the
Universe.
A small cosmological constant simply means that our position is far from the center. %cr

It is possible that the introduction of matter can connect the early
and late time accelerations with a decelerated era driven by matter.
However, to explain how this mechanism might happen is beyond the
scope of the present paper.

A potential problem in these type of models, is  that  objects
around us (like galaxies) look homogenously distributed. Analysis of
observational data suggests that we live in a homogeneous Universe
up to a scale of $\sim 140\ \mbox{Mpc}$ \cite{Hogg},\cite{Yadav}
(Below $100\ \mbox{Mpc}$, the inhomogeneities correspond to the
coarse graining of the energy density). This  constrains the allowed
inhomogeneity for a late-time mirage acceleration and does not rule
it out. It is also interesting to note that the observed
cosmological constant introduces a scale at which the
inhomogeneities of our model could in principle be observed. More
precisely, our model predicts inhomogeneities at scales just above
the observationally observable ones. It is intriguing that the very
small inhomogeneities of the CMB spectra from the WMAP data are
actually orthogonal to the ecliptic plane \cite{WMAP}. It has been
suggested that this phenomena can be explained introducing
inhomogeneities in the Universe \cite{Moffat}.
However, we need to test our model against the full CMB spectrum.
This could probably be done by considering the effective four-dimensional
Einstein equations obtained by re-writing the
geodesic equations of the brane in terms of standard General Relativity
with matter sources. In this case,
if these sources can be re-interpreted as coupled scalar fields, at least
under some approximations,
it might be possible to compare our model with standard cosmology
using the techniques developed in \cite{wands} for the
study of the CMB power spectrum and bispectrum on inhomogeneous background
geometries sourced by coupled scalar fields
\footnote{We thank the referee for pointing this out to us.}.
This is, however, postponed for future research.

Our model does not consider matter contributions.
In order to be
tested against BBN nucleosynthesys it must be implemented with
matter. Since early time matter and BBN constraints only deal with
radiation, one could imagine a consistent model introducing a small
contribution of Maxwell fields on the brane and then study
perturbations on it. We believe this will not spoil the acceleration
as the Maxwell fields will always be sub-dominant. However this have
to be tested with numerical calculations and is therefore beyond the
scope of this paper. An additional constraint would be to check our
model against the scale invariant perturbation observed in the CMB, although
it is not clear what a perturbation would mean in this case
as there is no homogeneous background. It should be noted that
another point, connected with the brane-matter contributions, which should be checked is the flatness of the Universe.
However, as long as we have not
introduce any matter we cannot answer this since an estimate of the  total
energy density versus the critical one is lacking in this case.

\ack

CGer and CGal would like to thank Misao Sasaki and Akihiro Ishibashi
for useful discussions about physical observers, and Ois{\'i}n Mac
Conamhna for comments on supergravity approximations. CGer would
like to thank Pierstefano Corasaniti and Bartjan VanTent for useful
discussions on observational cosmology. CGer~is supported by PPARC
research grant PPA/P/S/2002/00208. This work is co-funded by the
European Social Fund (75\%) and National Resources (25\%) -
(EPEAEK-B') -PYTHAGORAS.
%\maketitle

\end{document}